\documentclass[letter,twocolumn]{jpsj3}
\usepackage{txfonts}
\usepackage{graphics}
\usepackage{dcolumn}
\usepackage{color}
\usepackage{array,multirow,makecell}
\usepackage{bm}
\usepackage{nicefrac}
\usepackage{multicol,blindtext}
\usepackage{amssymb}

\newcommand{\UTe}{UTe$_2$}

\title{Field-reentrant superconductivity close to a metamagnetic transition in the heavy-fermion superconductor UTe$_2$}

\author{Georg \textsc{Knebel}$^{1}$, William \textsc{Knafo}$^2$, Alexandre \textsc{Pourret}$^{1}$, Qun \textsc{Niu}$^{1,3}$, Michal \textsc{Vali\v{s}ka}$^{1}$, Daniel \textsc{Braithwaite}$^{1}$, G\'{e}rard \textsc{Lapertot}$^{1}$,  Marc \textsc{Nardone}$^2$, Abdelaziz \textsc{Zitouni}$^2$, 
Sanu \textsc{Mishra}$^{3}$, Ilya \textsc{Sheikin}$^{3}$, Gabriel \textsc{Seyfarth}$^{3}$, Jean-Pascal \textsc{Brison}$^{1}$, Dai \textsc{Aoki}$^{1,4}$, Jacques \textsc{Flouquet}$^{1}$}

\inst{$^1$Univ. Grenoble Alpes, CEA, IRIG-Pheliqs, 38000 Grenoble, France  \\
$^2$Laboratoire National des Champs Magn\'{e}tiques Intenses, UPR 3228, CNRS-UPS-INSA-UGA, 143 Avenue de Rangueil, 31400 Toulouse, France\\
$^3$Laboratoire National des Champs Magnéetiques Intenses (LNCMI-EMFL), CNRS, Univ. Grenoble Alpes, 38042 Grenoble, France\\
$^4$Institute for Materials Research, Tohoku University, Ibaraki 311-1313, Japan\\
}

\abst{We present a study of the upper critical field of the newly discovered heavy fermion superconductor UTe$_2$ by magnetoresistivity measurements in pulsed magnetic fields up to 60~T and static magnetic fields up to 35~T. We show that superconductivity survives up to the metamagnetic transition at $H_{\rm m} \approx 35$~T at low temperature. Above $H_{\rm m}$ superconductivity is suppressed. At higher temperature superconductivity is enhanced under magnetic field leading to reentrance of superconductivity or an almost temperature independent increase of $H_{\rm c2}$. By studying the angular dependence of the upper critical field close to the $b$ axis (hard magnetization axis) we show that the maximum of the reentrant superconductivity temperature is depinned from the metamagnetic field. A key ingredient for the field-reinforcement of superconductivity on approaching $H_{\rm m}$ appears to be an immediate interplay with magnetic fluctuations and a possible Fermi-surface reconstruction. } 

\begin{document}
\maketitle

The discovery of coexistence of ferromagnetism and spin-triplet, equal-spin-pairing superconductivity (SC) in the orthorhombic  uranium compounds UGe$_2$ (under pressure),\cite{Saxena2000} URhGe\cite{Aoki2001} and UCoGe\cite{Huy2007}  (at ambient pressure) demonstrated directly that SC can be modified by tuning purposely the ferromagnetic fluctuations.\cite{Miyake2008, Wu2017}  Recent studies on these ferromagnetic superconductors have emphasized the interplay between ferromagnetic fluctuations and Fermi surface (FS) reconstructions associated with quantum phase transitions as a function of pressure $(p)$ or magnetic field $(H)$ (for a recent review see Ref.~\citen{Aoki2019a}).
 An illustrating example is URhGe: at $p=0$ ferromagnetic order occurs at $T_{\rm Curie} = 9.5$ K with an Ising sublattice magnetization $M_0 = 0.4 \mu_{\rm B}$ oriented along the $c$-axis of the orthorhombic crystal structure.\cite{Aoki2001} Applying a transverse magnetic field $H \parallel b$-axis leads to the suppression of $T_{\rm Curie}$ and to a spectacular reentrance of SC in the field range from 8~T to 13~T. \cite{Levy2005} At $H_{\rm R}=12$~T, a metamagnetic transition (MMT) occurs ($\Delta M_0 = 0.1 \mu_{\rm B}$) and the magnetic moments reorient from the $c$-axis ($H =0$) to the $b$-axis in the magnetic phase above $H_{\rm R}$. The enhancement of SC at $H_{\rm R}$ presumbly results from the combined effect of the enhancement of the ferromagnetic fluctuations \cite{Tokunaga2015, Kotegawa2015} and a Fermi-surface reconstruction.\cite{Yelland2011, Gourgout2016}. 
  
Similarly, the enhancement of the superconducting pairing under a transverse magnetic field appears in UCoGe, where the superconducting critical field $H_{\rm c2}$ along the $b$-axis shows an ``S''-shape driven by the collapse of $T_{\rm Curie}$.\cite{Aoki2009}  However, the ``S''-shape occurs in a field range far below the metamagnetic field $H_{\rm m} \approx 50$~T.\cite{Knafo2012}  The complete unraveling of the interplay between ferromagnetic fluctuations and topological FS changes suffers from the difficulty to detect major parts of the FS.  The discovery of new materials with similar properties is essential to trigger advances in the field.  

Recently, SC has been reported in the paramagnetic heavy fermion compound UTe$_2$.\cite{Ran2018, Aoki2019}  The superconducting transition temperature $T_{\rm sc} = 1.6$~K is remarkably high and the upper critical field $H_{\rm c2}$, exceeding by far the Pauli limitation for all crystallographic directions, has a very large anisotropy with an almost diverging $H_{\rm c2}$ for $H \parallel b$-axis at $T \approx 1$~K.  However, this upturn is found to be strongly sample dependent.\cite{Aoki2019}  UTe$_2$ has an orthorhombic crystal structure with $Immm$ space group. The easy magnetization axis is the $a$-axis, and the $c$-axis is initially the hard axis above 20~K. For $H \parallel b$, a peculiarity is the maximum of the magnetic susceptibility at $T_{\chi_{\rm max}} \approx 35$~K, so that at $T=2$~K the susceptibility is the lowest for the $b$-axis.\cite{Ikeda2006} This maximum of the susceptibility is linked to a MMT at $H_{\rm m} \approx 35$~T which has been observed in recent high field magnetization\cite{Miyake2019} and resistivity experiments\cite{Knafo2019} (see also Ref.~\cite{Aoki2013}). 

In the present paper, we concentrate on the superconducting phase diagram of different UTe$_2$ single crystals under high magnetic field along the $b$-axis. We  show that SC is strongly enhanced on approaching the MMT at  $H_{\rm m} \approx 35$~T. By turning the magnetic field away from the $b$-axis, the field-induced enhancement of SC is rapidly suppressed and  a usual orbital limited upper critical field is observed at an angle of 8 deg from $b$ to $a$-axis. 

Single crystals of UTe$_2$ were grown by chemical vapor transport with iodine as transport agent. We investigated three different single crystals from the same batch. They have been characterized by specific heat measurements and they all show a sharp superconducting transition at $T_{\rm sc} = 1.5$~K. The orientation of the crystals has been verified by Laue diffraction. The temperature dependence of $H_{\rm c2}$ has been first measured by resistivity in CEA Grenoble using a home-made dilution refrigerator with a base temperature of 100~mK and a superconducting magnet with field up to 16~T to measure  from 1.5~K downwards.  
Two of the measured crystals  (\#16 and \#12) were cut for high field measurements in Toulouse and Grenoble. 
In LNCMI-Toulouse we used a home-made dilution refrigerator at temperatures down to 200~mK in long duration (50~ms raise, 325~ms fall) pulsed magnetic field up to 58~T (samples \#16B and \#12A).  In  LNCMI Grenoble a top-loading dilution refrigerator with a Swedish rotator (with angular resolution of 0.1 deg) was operated down to 25~mK in magnetic fields up to 35~T (samples \#16D and \#12C). Sample \#16D has been turned in the $b-a$ plane and sample \#12C in the $b-c$ plane.  The resistivity measurements have been performed using a standard four point lock-in technique with current along the $a$-axis. Furthermore, we performed dc resistivity measurements in LNCMI Grenoble using a $^3$He cryostat up to 29~T on sample \#01. 

\begin{figure}
\begin{center}
\includegraphics[width=0.8\linewidth]{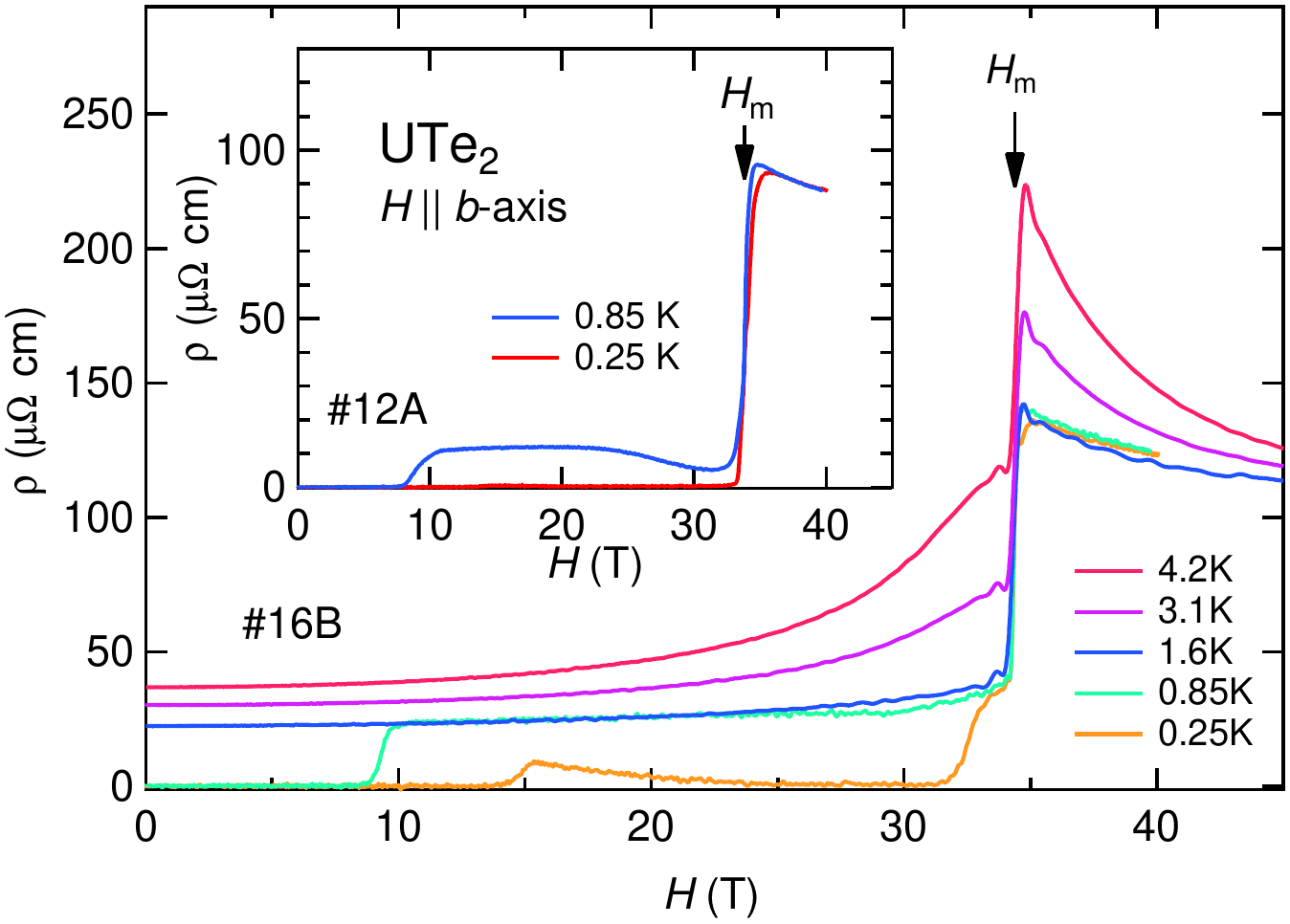}
\caption{Magnetoresistivity $vs.$ magnetic field $H$ along the $b$-axis of \UTe2 measured on sample \#16B and sample \#12A (inset) at different temperatures in pulsed magnetic field (down sweep).}
\label{f1}
\end{center}
\end{figure}

Figure \ref{f1} shows the magnetoresistivity of \UTe\ as a function of pulsed magnetic field applied along the $b$-axis at different temperatures for samples \#16B and \#12A. We only show the field-down sweeps, where the influence of eddy currents on the sample temperature are smaller due to the larger fall-time of the pulse (325~ms). At the lowest temperature ($\approx 250$~mK), in sample \#16B, $\rho =0$ up to 14.2 T. For higher fields the resistivity increases up to 37\% of the normal state value, but decreases again above 16~T all the way up to 24.5~T, where $\rho =0$ is observed. At 32~T, below the field of the MMT at $H_{\rm m} = 34.4$~T (defined by the maximum of the derivative $\partial \rho/\partial H$ from field sweep downwards), the $\rho (H)$ increases up to its normal state value. A huge step-like increase by a factor of four manifests the MMT with strong hysteresis (see Fig.~S1, supplemental material).
At 0.85~K shallow traces of this reentrant behavior of SC are still visible in the resistivity near 29~T. If we now look on sample \#12A (inset of Fig.~\ref{f1}), a very similar behavior is observed. At the lowest temperature, the resistivity is zero up to 33~T. A step-like increase of the resistivity at $H_{\rm m} = 33.9$~T indicates the MMT. At 0.85~K, the reentrance of SC shows up more strongly in the resistivity in this sample compared to sample \#16B. But again, slightly below the metamagnetic transition SC disappears abruptly. Just above $T_{\rm sc}$ at $T=1.6$~K, the $\rho (H)$ increases monotonously up to the step-like anomaly at $H_{\rm m}$ and decreases for $H > H_{\rm m}$. The transition field $H_{\rm m}$ is almost  temperature independent up to 4 K within our resolution, in agreement with a critical end point near $7-10$~K.\cite{Miyake2019, Knafo2019} The increase of $\rho (H)$ on both sides of $H_{\rm m}$ indicates a strong enhancement of the inelastic scattering term in the resistivity due to magnetic fluctuations and a possible underlying FS change. In Ref.~\citen{Knafo2019} we have shown that the $A$-coefficient of the Fermi-liquid resistivity $\rho (T) = \rho_0 +AT^2$ is enhanced by a factor of 6 at $H_{\rm m}$ compared to the zero field value. Assuming the validity of the Kadowaki-Woods ratio, this would suggest an increase of the effective mass $m^\star$ by a factor of 2.5 at $H_{\rm m}$. This enhancement of  $m^\star$ is in fairly good agreement with the derivation of $m^\star$ from magnetization measurements.\cite{Miyake2019} Importantly, no SC is observed above $H_{\rm m}$, possibly due to the new electronic state associated with a suddenly increased magnetization by a jump $\Delta M = 0.6\,\mu_{\rm B}$ and probably a strong change in the carrier number.

\begin{figure}
\begin{center}
\includegraphics[width=0.8\linewidth]{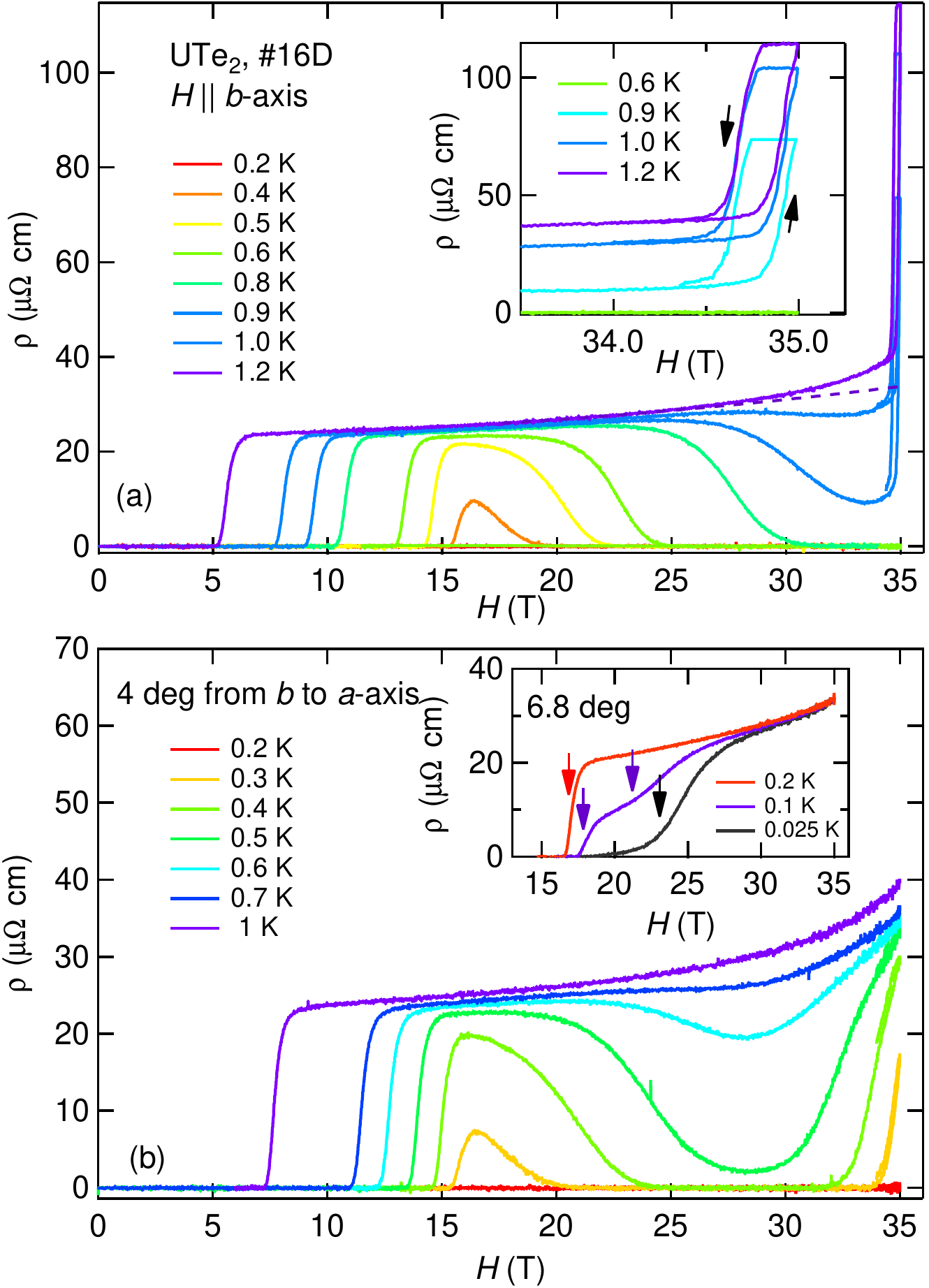}
\caption{(a) Magnetoresistance $vs.$ field of UTe$_2$ for $H \parallel b$ at different temperatures up to 1.2~K (sample \#16D). The dashed line is a fit with a usual $H^2$ field dependence of the magnetoresistivity at 1.2~K in the field range up to 26~T.  (b) $\rho (H)$ at an angle of 4~deg from the $b$ to $a$-axis.  The inset shows $\rho (H)$ at an angle of 6.8~deg from the $b$ axis at 0.025~K, 0.1~K and 0.2~K. The arrows indicate the superconducting critical fields (see text). }
\label{f2}
\end{center}
\end{figure}

To get a more detailed understanding of the superconducting phase diagram we performed resistivity measurements using a resistive magnet allowing steady fields up to 35~T. We aligned accurately the crystal \#16D along the $b$ axis by optimizing the sharp superconducting transition as a function of field at 0.6~K for different angles. The magnetoresistivity for $H \parallel b$-axis is shown in Fig.~\ref{f2} (a) for different temperatures. At 0.2~K the sample is superconducting up to the highest field of 35~T. At 0.4 K, $\rho (H)$ increases  between 15.4~T and 16.35~T up to a maximum value about 37~\% of the normal state resisitivity at this field. Such  an increase is similar to that in the pulsed field experiments on sample \#16B at at 0.25~K (see Fig.~\ref{f1}). For higher fields $\rho (H)$ decreases again and vanishes above 19.5~T. For higher temperatures the critical field of the onset of non-zero resistivity decreases while the appearance of reentrant SC shifts to higher fields. Above 0.8~K the resistivity does not vanish completely in the reentrance superconducting regime. Only a small decrease of $\rho (H)$ near 34~T can be followed up to 1~K. At 1.2~K no trace of reentrant SC appears and the normal state resistivity can be well fitted by a $H^2$ dependence up to 26~T. For higher fields deviations from the $H^2$-dependence  appear close to $H_m$ [see dashed line in Fig.~\ref{f2}(a)]. The stronger field dependence of the resistivity indicates the enhancement of critical scattering  in agreement with the strong increase of the $A$-coefficient of the $T^2$ of the resistivity (see Fig.~\ref{f1} and Ref.~\citen{Knafo2019}).

The inset of Fig.~\ref{f2}(a) shows the field dependence close to the MMT in an expanded view. While at 0.6~K zero resistivity is observed up to 35~T, we observed at 0.9~K a strong increase of the resistivity above 34.8~T.  From the size of the resistivity step in Fig.~\ref{f1} we can conclude that the transition shown in the inset of Fig.~\ref{f2}(a) is not complete and thus the hysteresis is not fully developped. In the pulsed field data shown in Fig.~\ref{f1}, $H_{\rm m}\approx 34.35$~T was found from the down sweep. However, a clear indication of the first order nature of the transition is the hysteresis of 0.25~T between field sweeps up and down, which is by far larger than the experimental resolution. For increasing temperatures up to 1.2~K, the field and hysteresis of the MMT do not change. 

In Fig.~\ref{f2}(b) we show $\rho (H)$ measured at an angle of 4 deg from the $b$ axis to the $a$-axis.  Overall it is very similar to that for $H \parallel b$, but the superconducting critical field is slightly lower and the reentrant superconducting phase is shifted to lower fields. Already at 0.3~K we find the onset to the normal state, but the resistivity vanishes again at 20~T and stays zero up to 34~T, where it again starts to increase. We do not find any indication of the 1$^{\rm st}$ order transition at $H_{\rm m}$, which we expect to be shifted to higher fields as we turn the field away from the $b$ axis. (The inset of Fig. S4 of Suppl. Material shows that the onset of the MMT is shifted by  0.1~T for 2 deg from the $b$-axis.) This seems different to the phase diagram of URhGe where the reentrant superconducting phase is initially pinned to the reorientation field of the magnetic moments and vanishes at a quantum critical end point  at a field of 15~T at an angle of about 6 deg from the $b$ axis.\cite{Yelland2011, Nakamura2017, Aoki2019a} In the inset of Fig.~\ref{f2}(b) we show $\rho (H)$ for an angle of 6.8~deg from the $b$-axis. The reentrant SC has almost collapsed. At 0.1~K we only see a broad, two-step transition, while at 0.2~K a sharp transition at 16.6~T marks the onset of the normal state. 

\begin{figure}[tb]
\begin{center}
\includegraphics[width=0.8\linewidth]{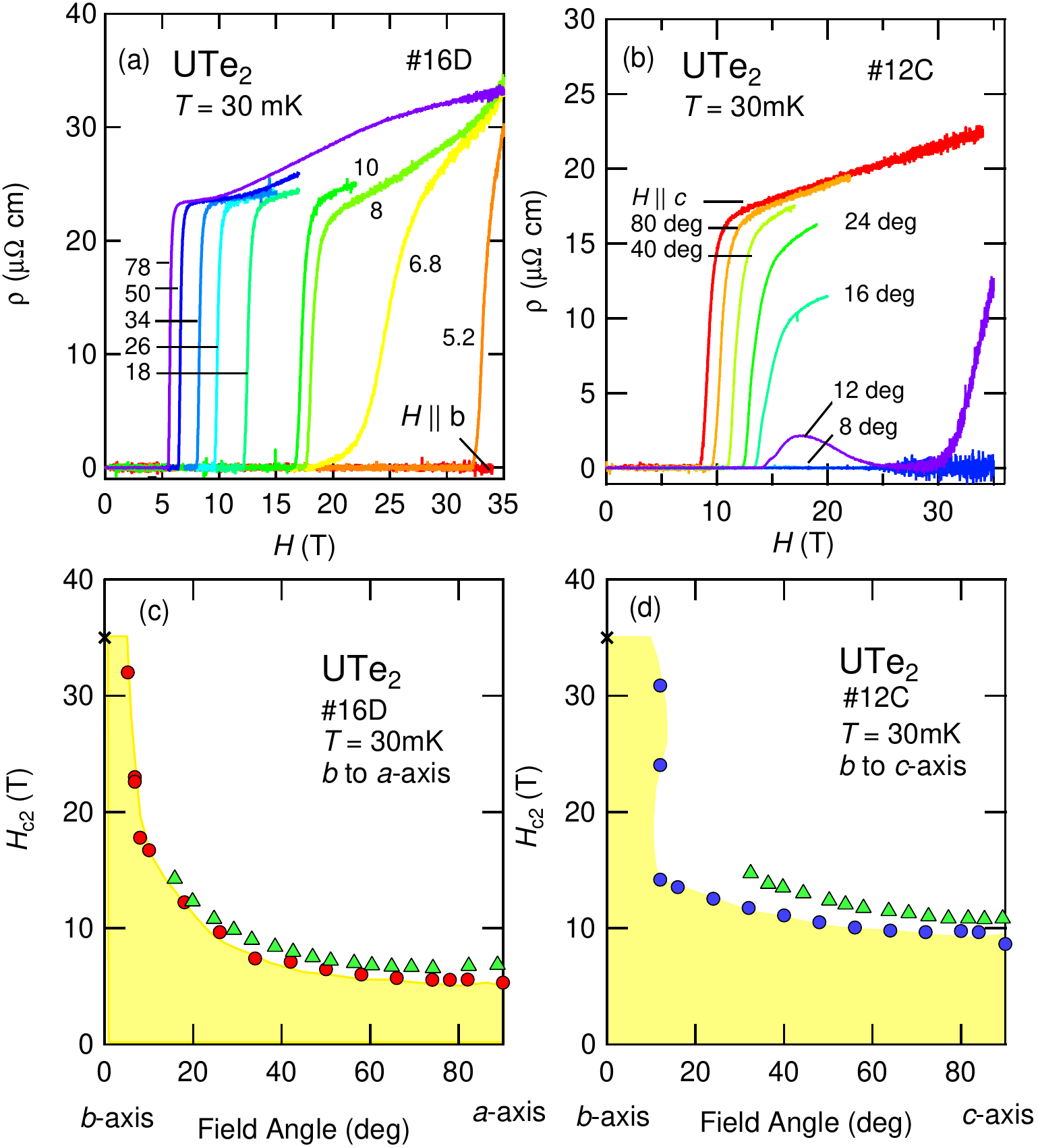}
\caption{Magnetoresistivity $vs.$ field of UTe$_2$ for different angles from (a) $b$- to $a$-axis and (b) from $b$- to $c$-axis at $T = 30$~mK (angles in deg from $b$). (c) and (d) show the corresponding angular dependencies of $H_{\rm c2}$ defined by zero resistivity, respectively. Crosses in (c) and (d) indicate $H_{\rm m}$. Green triangles in (c) and (d) show $H_{\rm c2}$ taken from Ref.~\citen{Aoki2019}. (We do not further discuss sample \#12C, as the sample quality degraded between the different experiments.)}
\label{f3}
\end{center}
\end{figure}
%


In Fig.~\ref{f3}(a) and (b) we show the angular dependence of $\rho (H)$ from $b$ to $a$-axis (sample \#16D) and from $b$ to $c$ axis (sample \#12C), respectively, at $T = 30$~mK. The anisotropy  of $H_{\rm c2}$ is shown in   Fig.~\ref{f3}(c) and (d) for both planes and it is compared to data published in Ref.~\citen{Aoki2019}. In the $(a,b)$ plane, SC is observed up to the highest field of 35~T up to an angle of 4~deg from the $b$-axis. By turning from the $b$-axis to the $c$-axis, SC  survives up to 35~T even for larger angles from the $b$-axis, in agreement with the anisotropy of the magnetic susceptibility ($\chi_a > \chi_c$). This very acute enhancement of $H_{\rm c2}$ near the $b$-axis cannot be explained by a conventional effective mass model assuming an ellipsoidal FS. This suggests that a more subtle mechanism  is responsible for the enhancement of SC near the $b$-axis. 


\begin{figure}[t]
\begin{center}
\includegraphics[width=0.8\linewidth]{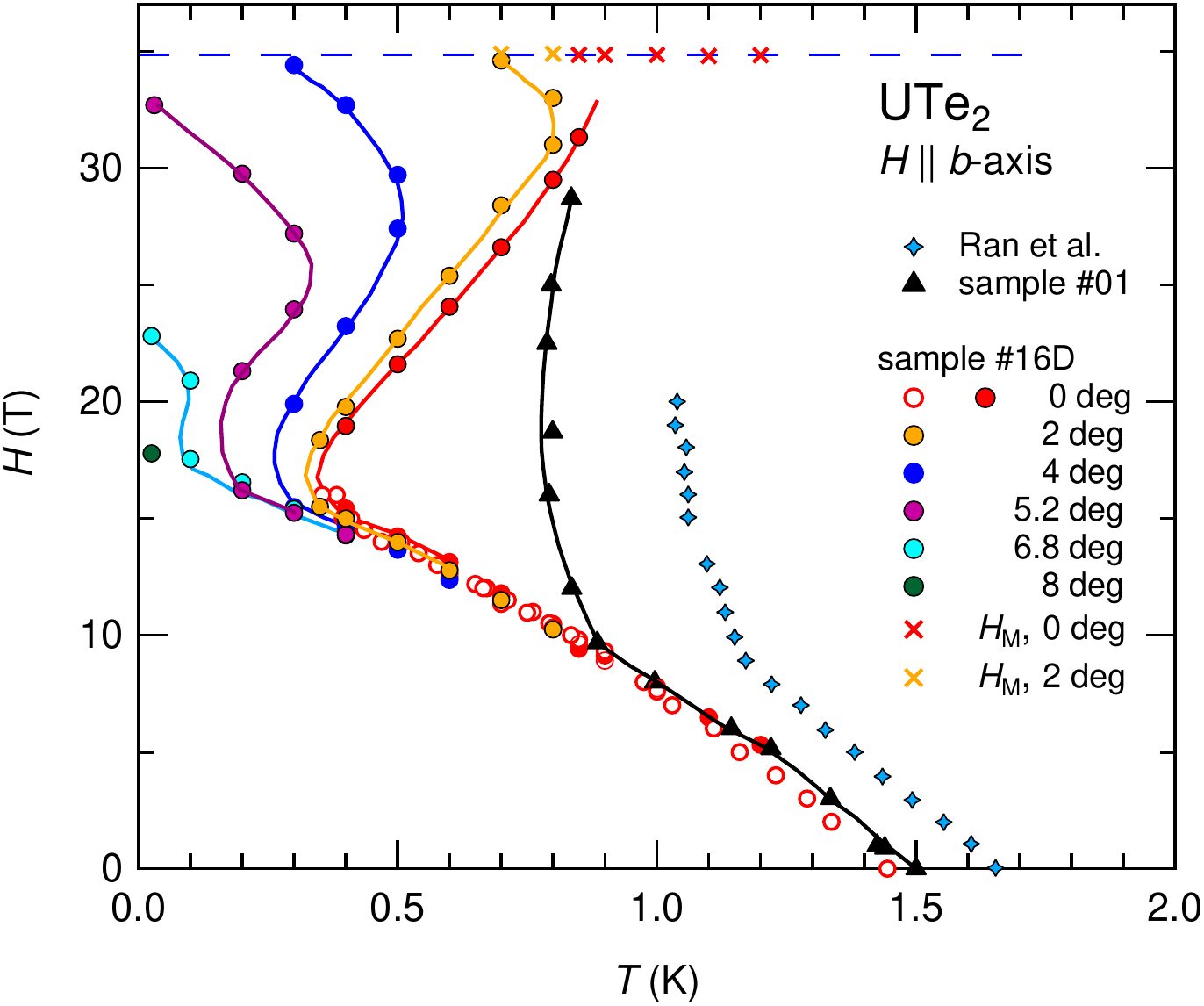}
\caption{Upper critical field $H_{\rm c2}$ of UTe$_2$ for different angles from the $b$- to the $a$-axis (circles). The superconducting transition is taken by the extrapolation $\rho \to 0$. Crosses give the metamagnetic field $H_{\rm m}$ for 0 deg and 2 deg from the $b$ axis. ($\rho (H)$ for all angles is shown in Fig.~S3 of Suppl. Material)  The dashed line indicates $H_{\rm m}$ at 0 deg. We also added $H_{\rm c2}$ from zero resistivity for sample \#01 (data see Fig. S2 in Suppl Material) and data taken from Ref.~\citen{Ran2018} (mid-point of the transition)  for comparison. Solid lines are guides to the eyes.}
\label{f5}
\end{center}
\end{figure}

\begin{figure}[t]
\begin{center}
\includegraphics[width=0.9\linewidth]{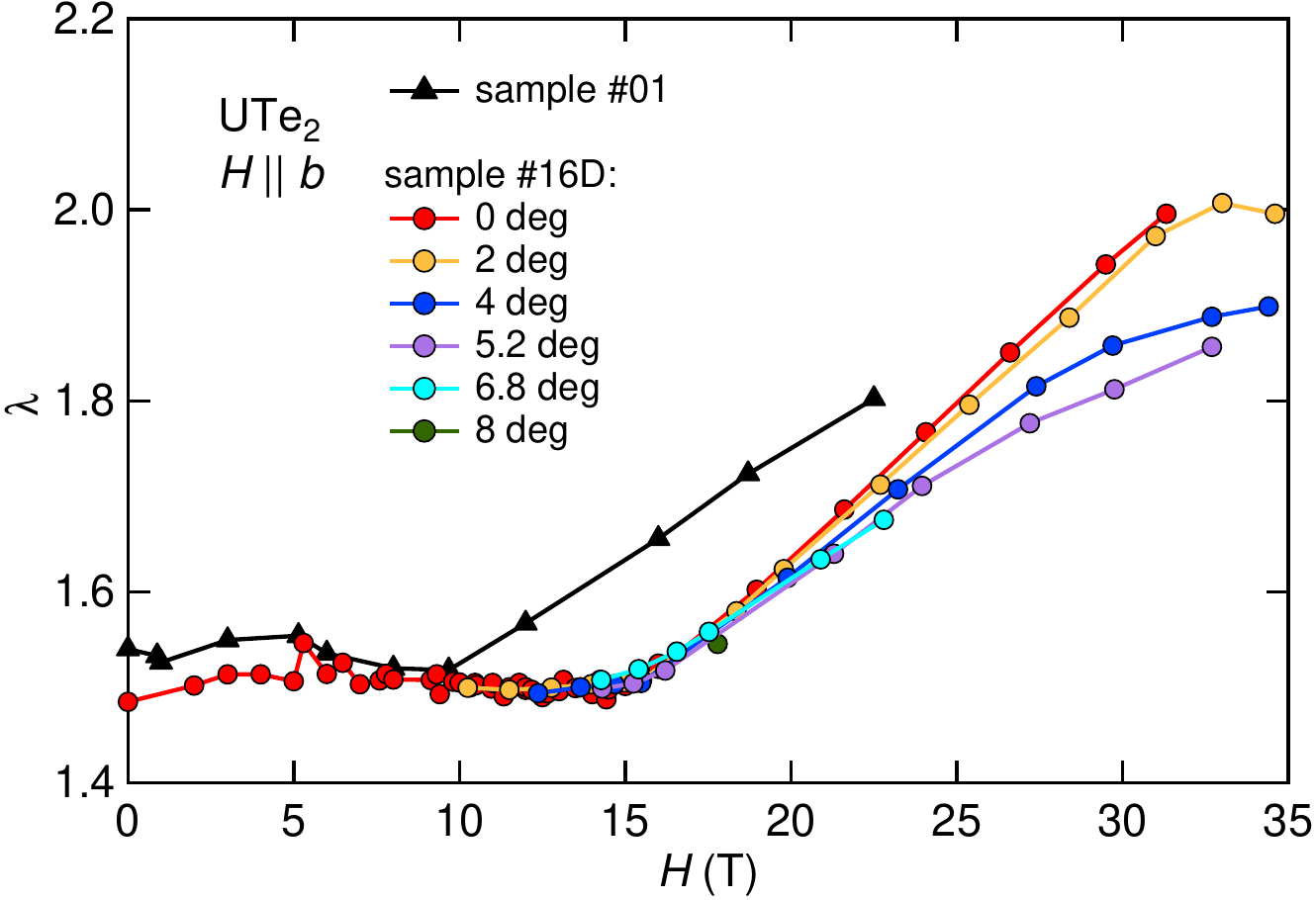}
\caption{Strong coupling parameter $\lambda (H)$ deduced from calculations of the upper critical field $H_{\rm c2}$ (see text). }
\label{f6}
\end{center}
\end{figure}

In Fig.~\ref{f5} we show $H_{\rm c2} (T)$ of UTe$_2$  in sample \#16D (circles) for $H \parallel b$-axis and for angles up to 8 deg turned from the $b$-axis to the $a$-axis. For comparison, we also added $H_{\rm c2}$ of sample \#01 and  the data taken from Ref.~\citen{Ran2018}. The initial temperature dependence of $H_{\rm c2}$ at low field of the different samples is very similar below 10~T, only the value of $T_{\rm sc}$ at zero field is different. $H_{\rm c2}$ from sample \#01 and and previous data in Ref.~\citen{Ran2018} show a quasi-divergence above 10~T. On the contrary  sample \#16D shows an almost usual $H_{\rm c2} (T)$ down to 300~mK and up to 15~T. For fields higher than 15~T a reentrance of SC appears and  $H_{\rm c2} (T)$  merges with that of sample \#01 above 29~T. For all samples the reentrant superconducting phase collapses at the MMT for $H \parallel b$. If we rotate the field from the $b$- to the $a$-axis, the reentrant SC is suppressed and it vanishes  near 17.8~T close to the orbital limited field for an angle of 8 deg.  
This result indicates that the superconducting pairing strength  can be tuned by the magnetic field in UTe$_2$.
There may be a combined effect of magnetic fluctuations associated to the MMT\cite{Knafo2019,Miyake2019} with a huge jump $\Delta M \approx 0.6\, \mu_{\rm B}$ and a possible FS reconstruction at $H_{\rm m}$.

In Fig.~\ref{f6} we show the field dependence of the strong coupling constant $\lambda (H)$ as extracted from calculations of $H_{\rm c2}$ for different (constant) values of $\lambda$, following what we did for UCoGe and URhGe.\cite{Wu2017}. The band Fermi velocity $v_{\rm F,band}$ has also been renormalised by the factor $(1+ \lambda (H))^{-1}$ and for the angular dependence, we corrected  $v_{\rm F,band}$ of its anisotropy between the $b$- and the $a$-axis:  we used an effective Fermi velocity $ \langle v_F\rangle _\theta$ controlling $H_{c2}$ at an angle $\theta$ from the $b$-axis: $\langle v_F\rangle _\theta = \left( \langle\ v_F\rangle _b^4 cos^2\theta +  \langle\ v_F\rangle_a^4 sin^2\theta \right)^{1/4}$, where $\langle\ v_F\rangle_b$ and $\langle v_F\rangle_a$ are extracted from the measured $H_{\rm c2}$ in these directions.
The main difficulty -- and also uncertainty -- is the adjustment of $v_{\rm F,band}$. In Ref.~\citen{Aoki2019} we had chosen a constant, isotropic $v_F$, following Ref.~\citen{Ran2018}, where $\langle v_F\rangle$ was determined from $H_{\rm c2}$ along $a$. This was implying a strong initial increase of $\lambda(H)$ for $H \parallel b$-axis, in order to reproduce the strong anisotropy of $H_{\rm c2}$ (a factor 3.2 at low fields in our samples). Here, we present another approach: we adjust $v_{\rm F,band} = 15000$~m/s for $H \parallel b$-axis on the measured initial slope at $T_{\rm sc}(0)$. Again, $v_{F,band}$ has been supposed to be field independent up to $H_{\rm m}$. With such parameters we find that $\lambda (H)$ is almost constant up to the field where the         upturn of  $H_{\rm c2}$ occurs i.e.~at 10~T and 15~T for samples \#01 and \#16D, respectively. This means that $H_{\rm c2}$ along b is well described by the orbital limit up to the upturn. For higher fields, $\lambda$ increases almost linearly from 1.5 to 2. At finite angles, the re-entrant phase is also described by a monotonous increase of $\lambda(H)$, and results solely from the tight competition between orbital limitation and field-increase of the coupling strength. We have chosen a larger value of $\lambda(0)=1.5$ than previously ($\lambda(0)=0.75$ in Ref.~\citen{Ran2018,Aoki2019}), because the field dependence of the $A$ coefficient or the $\gamma$ coefficient \cite{Knafo2019, Miyake2019} suggest a strong field-increase of $\lambda(H)$. Nevertheless, even with $\lambda(0)=1.5$, we could not reproduce the suggested strong enhancement close to $H_{\rm m}$. Clearly, we need more precise measurements of the field dependence of the Sommerfeld coefficient in order to cancel the arbitrariness of the choice of $v_{\rm F,band}$ and of $\lambda(0)$. The approach could be also over-simplified in the neighborhood of $H_{\rm m}$, where a significant evolution of  $v_{\rm F,band}$ could take place.

Indeed, above $H_{\rm m}$ SC disappears abruptly, (see Fig.\ref{f1}). The very large resistivity jump at $H_{\rm m}$ \cite{Knafo2019} strongly suggest a loss of carrier numbers above $H_{\rm m}$, and so a strong decrease of the density of states, as well as a FS reconstruction. If correlation effects are also suppressed due to the band polarisation \cite{Miyake2019}, we should also expect a large increase of $v_F$, hence, a drastic reduction of the orbital limit. All of these effects are expected to suppress SC above $H_{\rm m}$, but a FS evolution could already start to emerge before the MMT.

When the field direction is turned away from the $b$-axis, the MMT is expected to increase in field: this is naturally the case if, like in URhGe, it depends mainly on the projection of the induced magnetization along the $b$-axis. Similar to the case of URhGe, in UTe$_2$ the critical end point of the $H_{\rm m}$ line may depends drastically on the angle from the $b$-axis and the enhancement of $m^\star$ and thus of $\lambda$ should be suppressed. 
In URhGe the reentrant SC is pinned to the reorientation of the magnetic moment, which is further supported by the uniaxial stress dependence of $H_{\rm m}$.\cite{Braithwaite2018}  Here in UTe$_2$, the misalignment with respect to the  $b$-axis  leads to a depinning of the $H_{\rm c2}$ from $H_{\rm m}$.
The fact that we do not observe a marked enhancement of $\lambda (H)$ on approaching $H_{\rm m}$ is even more evident on the softening of the increase of $\lambda (H)$ with increasing angle from the $b$-axis, as shown in Fig.~\ref{f6}.

In heavy-fermion systems, a MMT is often accompanied by a FS instability.\cite{HAoki2014, Mydosh2017} The large jump of the $b$-axis magnetization $\Delta M \approx 0.6\, \mu_{\rm B}$ must have some feedback on the FS. 
Indeed it,is almost comparable to the magnetization jump in UGe$_2$ above $p_{\rm c}$ on entering in the weakly polarized phase FM1 under magnetic field.\cite{Pfleiderer2002} Here, the crossings of phase boundary from paramagnetic to FM1 is associated to a FS reconstructions. \cite{Terashima2001, Terashima2002} In URhGe a Lifshitz transition is coupled to the field reorientation of the magnetic moment and the reentrant SC.\cite{Yelland2011, Gourgout2016} It has been proposed that the multiband nature of this system may be responsible for the mass enhancement of one of the bands associated with the topological FS change.\cite{Sherkunov2018} 
Note that the field dependence of the specific heat at 0.4~K in UTe$_2$ shows a rapid  increase at low field, implying a multiband SC.\cite{Aoki2019}
The main difference with URhGe (and other ferromagnetic superconductors) is that UTe$_2$  is a paramagnet in the normal state, i.e.~down to $T_{\rm sc} =1.6$~K. Up to now, no ferromagnetic component has been detected below $T_{\rm sc}$. However, a non-unitary superconducting state with pairing of only one spin state has been proposed.\cite{Ran2018} In that case the superconducting transition would imply a secondary ferromagnetic order and a spontaneous spin splitting of the subbands which has not been established experimentally yet.

In conclusion we have studied the upper critical field in the heavy fermion superconductor UTe$_2$ for magnetic field along the hard magnetization $b$-axis. We showed that SC extends at low temperature up to the MMT at $H_{\rm m} \approx 35$~T, with a reentrance of SC on approaching $H_{\rm m}$.  Turning the magnetic field from the $b$-axis suppresses the reentrant superconducting phase, which is decoupled from $H_{\rm m}$. The temperature dependence of $H_{\rm c2}$ is explained by a balance between the orbital limiting field and the enhancement of SC due to the fluctuations associated with the MMT.

\begin{acknowledgment}
Part of this work at the LNCMI was supported by Programme Investissements d'Avenir under the program ANR-11-IDEX-0002-02, reference ANR-10-LABX-0037-NEXT. We acknowledge the financial support of the Cross-Disciplinary Program on Instrumentation and Detection of CEA, the French Alternative Energies and Atomic Energy Commission, the ANR-DFG grant "Fermi-NESt", and KAKENHI (JP15H05882, JP15H05884, JP15K21732, JP16H04006, JP15H05745, JP19H00646).

\end{acknowledgment}


\newpage
\vfill
\noindent
{\bf \large  Supplemental Material for\\
Field-reentrant superconductivity close to a metamagnetic transition in the heavy-fermion superconductor UTe$_2$}


\vspace{1cm}




In this Supplemental Material we show complementary data to those presented in the main article. 

%
%

\section{Magnetoresistivity in pulsed field}

\begin{figure}[h]
\makeatletter
\renewcommand{\thefigure}{S\@arabic\c@figure}
\makeatother
\begin{center}
\includegraphics[width=0.8\linewidth]{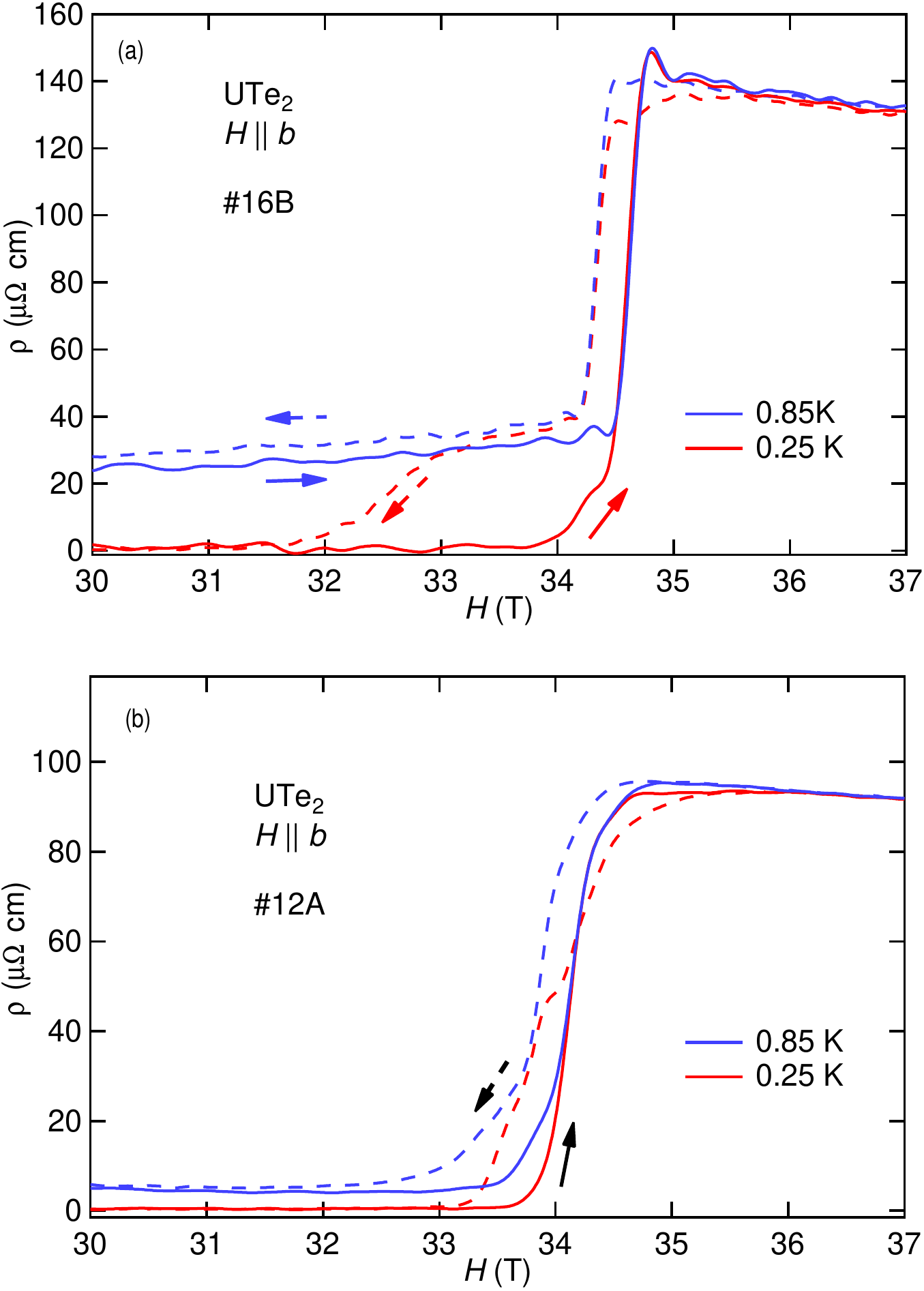}
\caption{Resistivity {\em vs.} field at for two different samples in pulsed field, solid lines for field sweep up, dashed lines for field sweep down.}
\label{toulouse}
\end{center}
\end{figure}

Figure~\ref{toulouse} compares field-up and field-down data obtained for samples \#16B and \#12A in pulsed magnetic fields in the vicinity of $H_{\rm m}$. A similar hysteresis, of field width $\simeq0.3$~T, is observed for both samples at the metamagnetic field $H_{\rm m}$. In sample \#16B, a small misalignment from the $b$-axis may be responsible for the observed higher value of  $H_{\rm m}$ and smaller value of $H_{\rm c2}$. We note that a hysteresis at $H_{\rm c2}$ is bigger than that at $H_{\rm m}$, possibly due to a deviation from non-isothermal conditions resulting from the use of pulsed magnetic fields (magnetocaloric and eddy currents effects).

\section{Magnetoresistivity of sample \#01}

\begin{figure}[t]
\makeatletter
\renewcommand{\thefigure}{S\@arabic\c@figure}
\makeatother
\begin{center}
\includegraphics[width=0.9\linewidth]{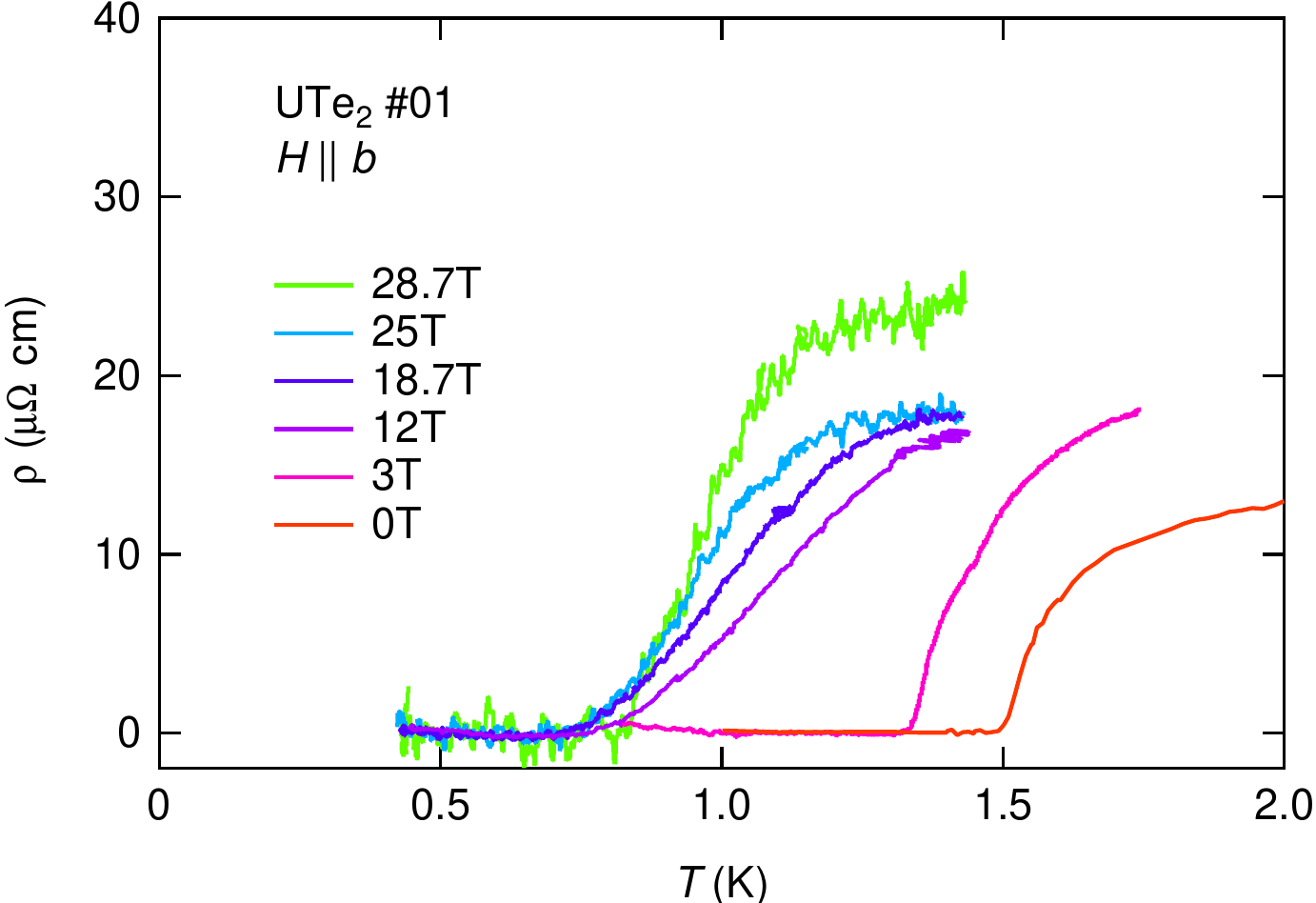}
\caption{Resistivity {\em vs.} temperature at various magnetic fields up to 28.7~T measured on sample \#01. Interestingly the superconducting transition gets sharper for high magnetic fields. }
\label{sample1}
\end{center}
\end{figure}

Previously, we have already recognized that the superconducting properties of UTe$_2$ are strongly sample dependent, especially for a field applied along the $b$-axis.\cite{Aoki2019} Ran et al.~\cite{Ran2018} reported a strong upturn of $H_{\rm c2}$ to occur already for $T \approx 1$~K. In Fig.~\ref{sample1} we show the temperature dependence of the resistivity of sample \#01 measured in static fields up to 28.7~T.  Obviously, $H_{\rm c2}$, defined by $\rho = 0$, is almost temperature independent for $H > 12$~T up to 28.7~T and the behavior of this sample resembles very much to that reported in Ref.~\citen{Ran2018}. Interestingly the transition width decreases with increasing magnetic field indicating the strengthening of superconductivity on approaching $H_{\rm m}$. The varying behavior of different samples was suspected to depend on the initial $T_{\rm sc} (0)$.\cite{Aoki2019}

\section{Angular dependence of the magnetoresistivity}

\begin{figure}[t]
\makeatletter
\renewcommand{\thefigure}{S\@arabic\c@figure}
\makeatother
\begin{center}

\includegraphics[width=0.9\linewidth]{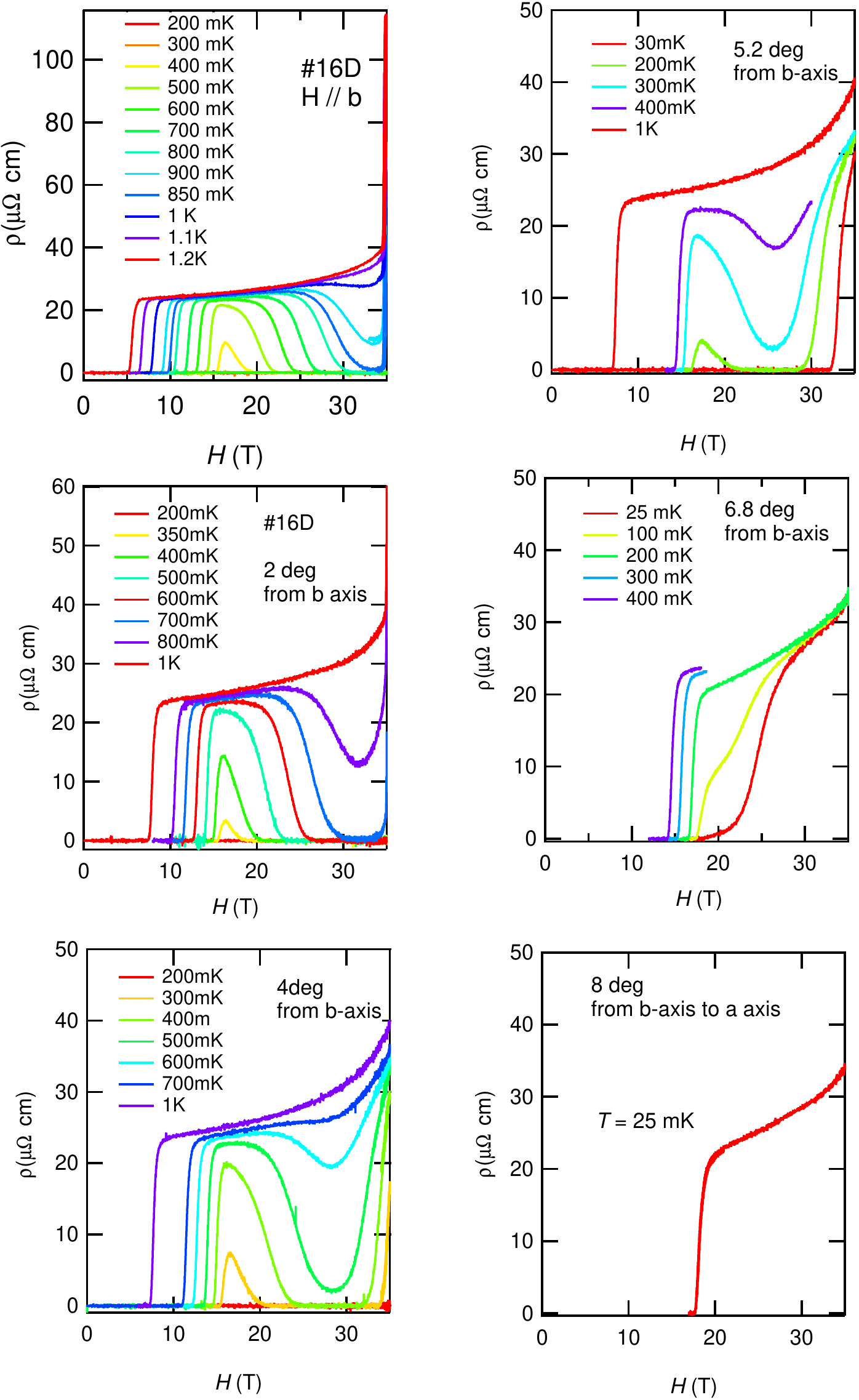}
\caption{Magnetoresistivity {\em vs.} field of UTe$_2$ at different temperature for various angles measured on sample \#16D. }
\label{supplf2}
\end{center}
\end{figure}

\begin{figure}[b]
\makeatletter
\renewcommand{\thefigure}{S\@arabic\c@figure}
\makeatother
\begin{center}

\includegraphics[width=0.9\linewidth]{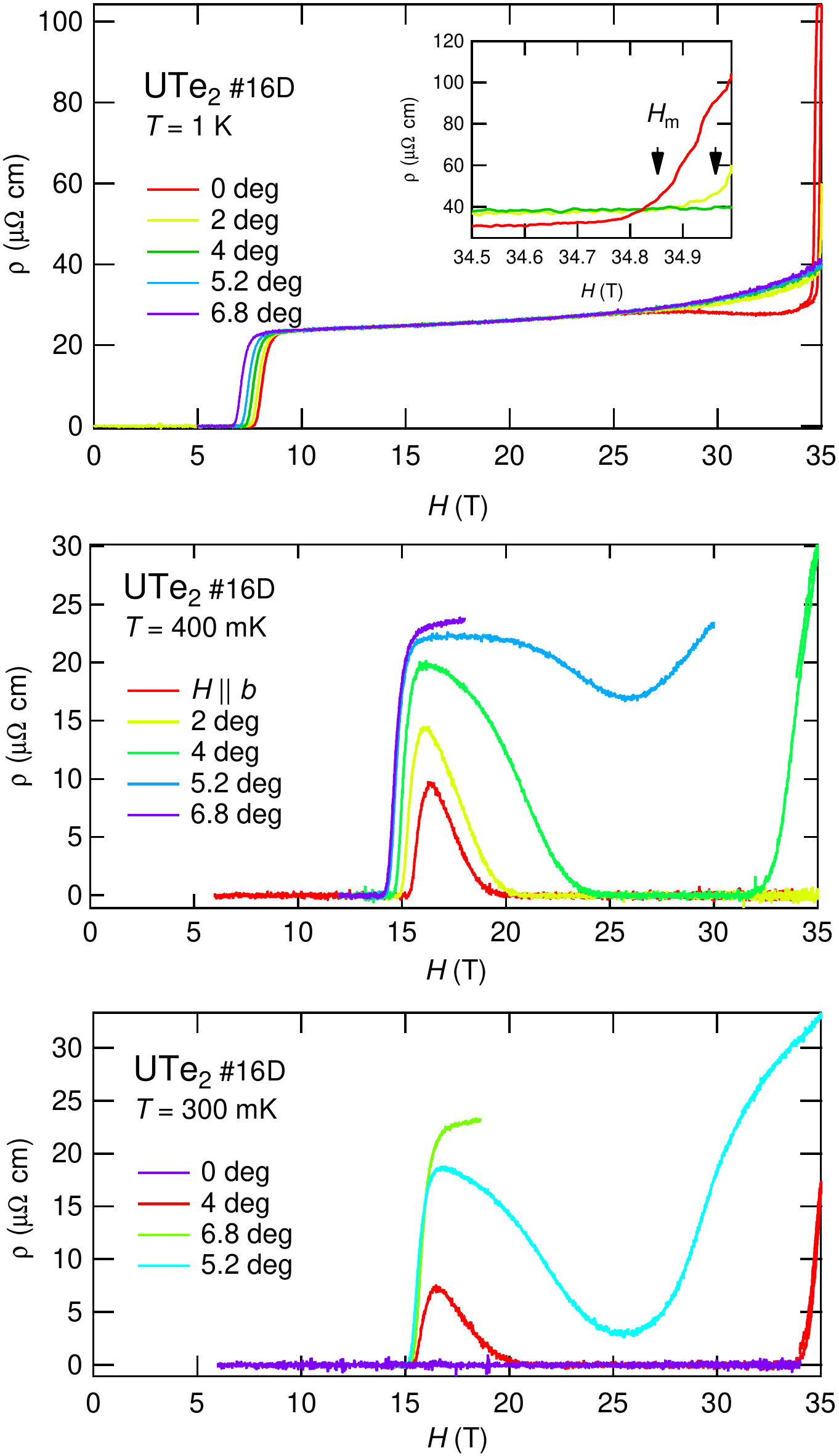}
\caption{Comparison of the magnetoresistivity {\em vs.} field of UTe$_2$ at fixed temperature different  for various angles measured on sample \#16D. The inset in the upper panel shows a zoom on the metamagnetic transition for $H \parallel b$, 2 deg, and 4 deg. The arrows indicate $H_{\rm m}$. }
\label{supplf3}
\end{center}
\end{figure}

Figure~\ref{supplf2} shows the magnetoresistivity {\em vs.} field of UTe$_2$ at different temperature for various angles from the $b$-axis measured on sample \#16D. The data clearly show that the reentrant superconductivity at $H \parallel b$ and at 2~deg collapses at the metamagnetic transition. For higher angles this transition shifts out of our accessible field window above 35~T (see inset of Fig.~\ref{supplf3}). However, it is obvious that the reentrant superconductivity is suppressed at lower fields for  higher field angles. At 8 deg, the reentrant superconducting phase is fully suppressed. 

In Fig.~\ref{supplf3} we show the magnetoresistivity {\em vs.} field of UTe$_2$ at $T=1$~K (upper panel), $T=0.4$~K (middle panel) and $T = 0.3$~K (lower panel). Obviously, with increasing angle the reentrant superconductivity detaches from the metamagnetic field, which can be followed only up to 2 deg from the $b$-axis in the field window up to 35~T as shown in the inset of Fig.~\ref{supplf3}. The shallow minimum for fields just below 35~T at $T=1$~K for $H\parallel b$ is due to the reentrant superconductivity. The field dependence of the magnetoresistivity at lower temperatures ($T=400$~mK and $T=300$~mK) supports that that the enhancement of superconductivity diminishes   away from the $b$ axis. 

\section{Determination of the critical field} 

\begin{figure}
\makeatletter
\renewcommand{\thefigure}{S\@arabic\c@figure}
\makeatother
\begin{center}
\includegraphics[width=0.9\linewidth]{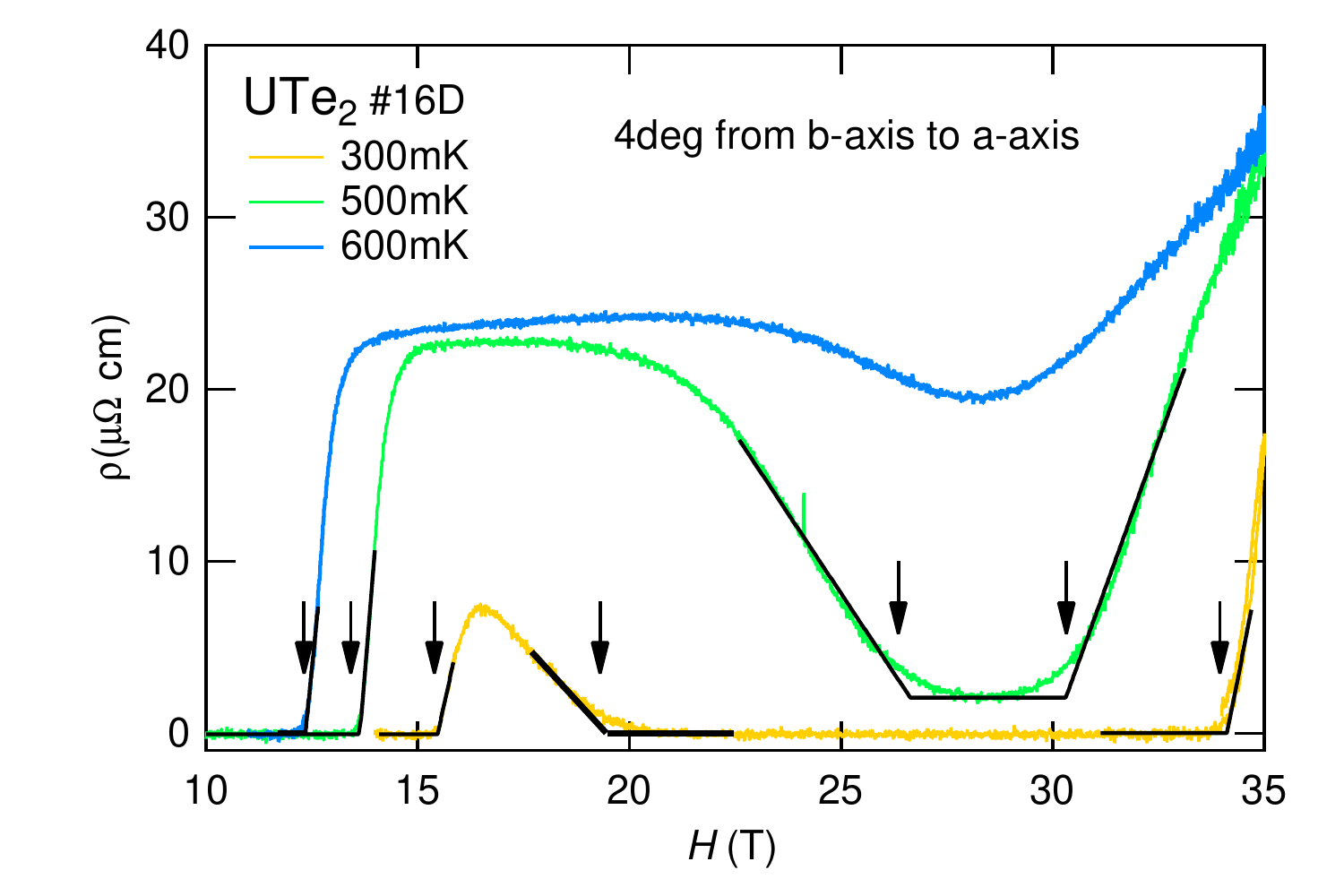}
\caption{Example of the determination of the critical temperatures of the superconducting transition. Arrows indicate the transition fields reported in Fig.~4 of the main paper. }
\label{supplf4}
\end{center}
\end{figure}

Figure \ref{supplf4} gives an example of the determination of the transition temperature reported in Fig.~4 of the main paper. The magnetoresistivity is shown for an angle of 4 deg from the $b$-axis to the $a$-axis for different temperatures. While $\rho (H)$ vanishes in the reentrant phase at 300~mK, the resistivity does not fall to zero in the field range fron 26~T to 30~T at 500~mK. However, in that range the lowest resitivity is less than 10\% of the normal state resistivity at that field and we fitted the data as shown in Fig.~\ref{supplf4}. In difference, for $T=600$~mK, the drop of the resistivity is only down to 75\% of the normal state value and thus we did not report these point in the phase diagram of Fig.~4 of the main paper. We choose the criteria that $\rho$ should at least be less than 30\% of the normal state  resisitivity. 
critical fields have been determined by fitting two straight lines to the data as shown in the figure.

\section{Determination of $\lambda (H)$}

\begin{figure}
\makeatletter
\renewcommand{\thefigure}{S\@arabic\c@figure}
\makeatother
\begin{center}

\includegraphics[width=0.9\linewidth]{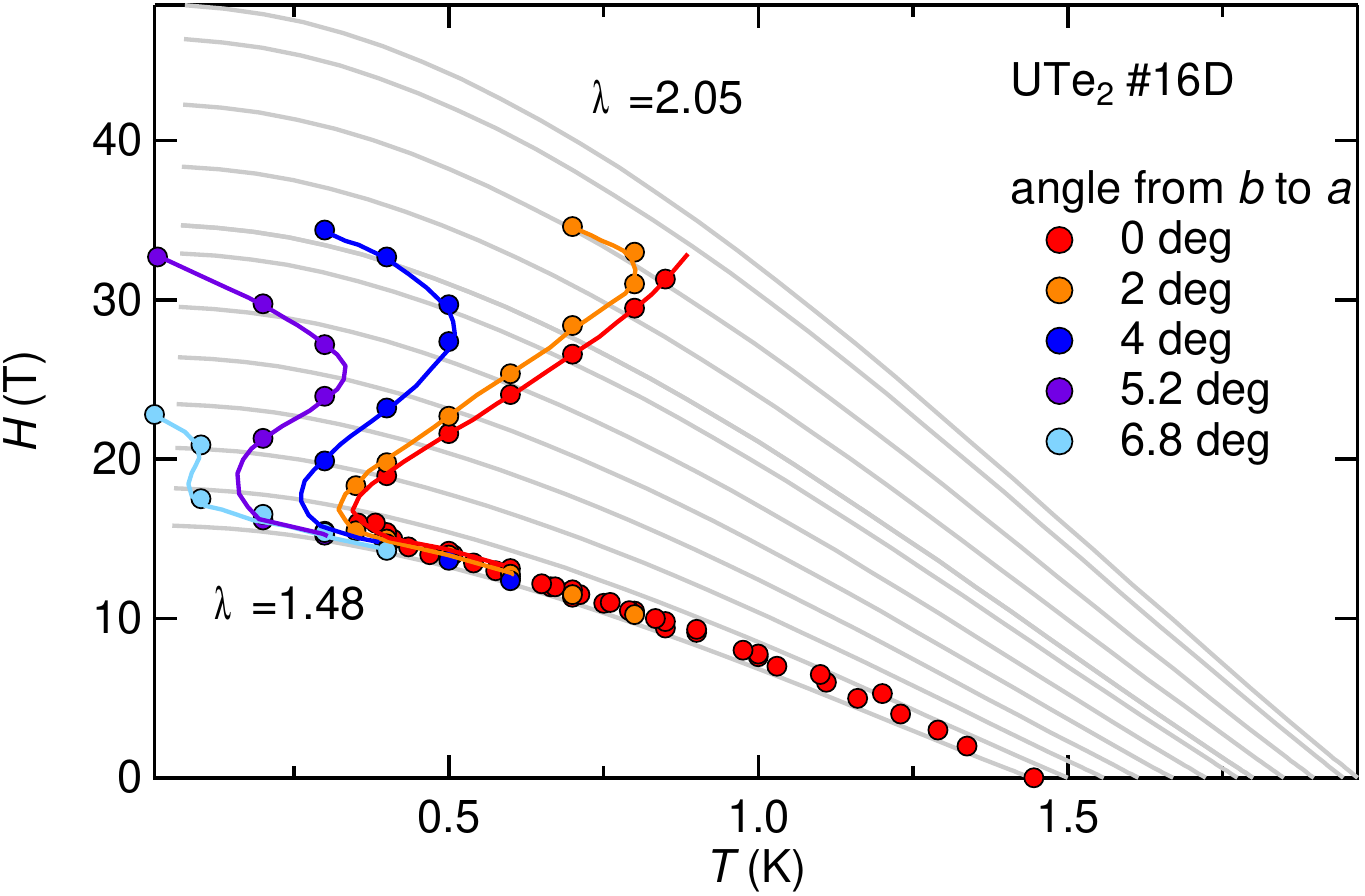}
\caption{Determination of $\lambda (H)$. Gray lines are $H_{\rm c2}$ calculations for $\lambda$ varied between 1.48 and 2.05 by steps of 0.06 used to extract $\lambda (H)$. }
\label{supplf5}
\end{center}
\end{figure}

Figure~\ref{supplf5} shows, how $\lambda (H)$ has been determined. Details of the calculations are explained in Ref.~\citen{Wu2017}. Gray lines show calculations of $H_{\rm c2}$  for $\lambda$ varied between 1.48 and 2.05 by steps of 0.06. 
 \vspace{3cm}

\end{document}